\documentclass[12pt]{iopart}          
\usepackage{url}

\RequirePackage[T1]{fontenc}

\RequirePackage{multirow}
\RequirePackage{graphicx}
\RequirePackage{flushend}
\RequirePackage[colorlinks,citecolor=blue,urlcolor=blue,linkcolor=blue]{hyperref}

\usepackage{graphicx}  
\usepackage{dcolumn}   
\usepackage{bm}        
\usepackage{amssymb}   
\usepackage{feynmf}    
\usepackage{hyperref}  %
\usepackage{slashed}
\usepackage{color}
\usepackage{array}
\usepackage{caption}
\usepackage{lineno}
\usepackage{subcaption}
\expandafter\let\csname equation*\endcsname\relax
\expandafter\let\csname endequation*\endcsname\relax
\usepackage{amsmath}
\usepackage{booktabs}
\usepackage{multirow}
\usepackage{subcaption}
\usepackage{wrapfig}
\usepackage{dblfloatfix}
\usepackage{url}
\usepackage{relsize}
\usepackage[style=numeric-comp,backend=biber,sorting=none,maxcitenames=4, mincitenames=4,
maxbibnames=99, minbibnames=99]{biblatex}
\usepackage[shortlabels]{enumitem}

\usepackage{parskip}

\newsavebox\mybox

\begin{document}

\title[Accelerating Graph-based Tracking]{Accelerating Graph-based Tracking Tasks with Symbolic Regression}

\author{Nathalie Soybelman$^1$, Carlo Schiavi$^2$, Francesco A. Di Bello$^2$, Eilam Gross$^1$}
\address{$^1$ Weizmann Institute of Science, Israel}
\address{$^2$ University of Genova and INFN Sezione di Genova, Italy}

\ead{nathalie.soybelman@weizmann.ac.il}
\vspace{10pt}
\begin{indented}
\item[]April 2024
\end{indented}

\begin{abstract}
The reconstruction of particle tracks from hits in tracking detectors is a computationally intensive task due to the large combinatorics of detector signals. Recent efforts have proven that ML techniques can be successfully applied to the tracking problem, extending and improving the conventional methods based on feature engineering. 
However, complex models can be challenging to implement on heterogeneous trigger systems, integrating architectures such as FPGAs.
Deploying the network on an FPGA is feasible but challenging and limited by its resources. An efficient alternative can employ symbolic regression (SR). We propose a novel approach that uses SR to replace a graph-based neural network. Substituting each network block with a symbolic function preserves the graph structure of the data and enables message passing. 
The technique is perfectly suitable for heterogeneous hardware, as it can be implemented more easily on FPGAs and grants faster execution times on CPU with respect to conventional methods.
While the tracking problem is the target for this work, it also provides a proof-of-principle for the method that can be applied to many use cases.
\end{abstract}
\vspace{1pc}
\noindent{\it Keywords}: graph neural networks, object condensation, symbolic regression
\vspace{1pc}

\section{Introduction}
In High Energy Physics experiments, such as ATLAS~\cite{ATLAS} and CMS~\cite{CMS} at the LHC~\cite{LHC}, data collection relies on sophisticated trigger systems~\cite{ATLAS_trigger,ATLAS_trigger_run3,cms_trigger,cms_trigger_run3} that select interesting events and reject others to significantly reduce the data rate from $\mathcal{O}$(40 MHz) to $\mathcal{O}$(1 kHz) for final storage. With the upcoming High Luminosity LHC upgrades, the instantaneous luminosity will increase from 1 to $7.5 \times 10^{34}\, \text{cm}^{-2} \text{s}^{-1}$, and the number of simultaneous collisions per bunch crossing (pile-up) will rise from 20-50 to 140-200, reaching unprecedented heights. These changes necessitate substantial improvements in data acquisition pipelines to handle the increased data rates efficiently. In this context, the inner tracking detector data, which captures hits from charged particles, plays a vital role in complementing calorimeter and muon detector data for triggering, enabling more precise particle reconstruction and identification. However, track reconstruction from these hits is challenging due to the combinatorics of detector signals, making it one of the most time-consuming trigger tasks~\cite{ATLAS_trigger_tracking}.

Numerous machine learning (ML) algorithms have been explored to improve upon classical track reconstruction approaches, particularly in the context of the \textit{TrackML} challenge~\cite{trackml-particle-identification,trackml-sum} and beyond. The most promising methods are based on graph-neural networks~\cite{GNN_general,gnn_ckf,gnn_edgeclass,gnn_equivariant,gnn_hierarchical,gnn_hllhc,gnn_Thais}, which provide a natural way of representing and processing tracking data. 

However, full deployment of these methods - characterized by large memory requirements and non-negligible execution times - into the ATLAS and CMS trigger strategies remains an open challenge. Currently, both trigger systems perform tracking only in the software-based high-level trigger (HLT), which still primarily relies on CPU farms. CMS has recently introduced GPUs to offload part of the tracking task~\cite{CMS_trigger_gpu_tracking}, though their potential is not yet fully exploited. At the same time, it is essential to explore alternative approaches based on heterogeneous hardware solutions, such as adopting Field Programmable Gate Arrays (FPGAs). 

Speeding up the inference of ML tools through quantization and FPGA implementation~\cite{fpga_generic,fpga_hls4ml,fast_muon_fpga,fpga_convnet} could extend tracking capabilities to the hardware-based early stages of trigger selection~\cite{HW-tracking}. This approach can incorporate GNNs or transformer architectures, as demonstrated in~\cite{GNN_fpga,gnn_fpga2,gnn_fpga_reco,fpga_traf}. However, it is only feasible for relatively small and simple networks (on the order of 10k parameters), as large-scale GNNs face memory limitations. 

Symbolic regression can replace or approximate dense neural networks with analytical functions. Previous studies~\cite{SR_first, SR_BSM, SR_HI} have employed this technique to derive more interpretable and physics-motivated expressions. Moreover, it can accelerate inference and ease FPGA implementation without significant performance loss, as explored in~\cite{fpga_sr}. 

In this work, we propose a novel extension of this method, applying symbolic regression to approximate a graph neural network for track finding. This task is challenging because the network input is not a simple list of variables but a set of objects with varying properties and no fixed cardinality. Thus, finding a single analytical function to map hits directly to tracks is unfeasible. Instead, we decompose the network into its dense layers and fit them individually with symbolic expressions. This approach preserves the graph structure and message-passing mechanism, offering flexibility: the entire GNN or a subset of its components can be replaced with symbolic expressions, while the rest remains as neural networks. 

To test this approach, we developed an algorithm to cluster hits originating from high transverse momentum tracks, separating them from hits produced by soft background tracks and detector noise. The clustered signal hits can be processed further to extract track parameters via a fit procedure.

For a proof-of-concept, we designed a simplified toy data simulator, enabling a direct comparison between the performance of the clustering GNN and the symbolic regression approximations.

\section{Dataset}
The toy data generator used in this study simulates a cylindrical detector consisting of 8 concentric layers. Its geometry is based on the barrel of the ATLAS Inner Detector, with layer radii matching those of the Pixel and SCT (SemiConductor Tracker) subdetectors, each composed of 4 layers. 
The emulated detector operates within a constant 2~T solenoidal magnetic field aligned with the beam axis, approximating the setup of the ATLAS experiment.

The generator tracks the trajectories of simulated charged particles in the detector volume and calculates their intersection with the detector layers to generate hits. To maintain simplicity, no particle interactions with the detector material are considered.

In this study, we adopt the same coordinate system used by ATLAS and CMS\footnote{We adopt a right-handed coordinate system with the origin at the nominal interaction point (IP) in the centre of the detector, with the \(z\)-axis along the beam pipe. The \(x\)-axis points from the IP to the centre of the accelerator ring, and the \(y\)-axis points upwards. Polar coordinates \((r,\phi)\) are used in the transverse plane, where \(\phi\) is the azimuthal angle around the \(z\)-axis. The pseudorapidity is defined in terms of the polar angle \(\theta\) as \(\eta = -\ln \tan(\theta/2)\).}.

The $\phi$ and $z$ coordinates of the hits are smeared based on the pitch of the emulated detector sensors to account for experimental resolution effects, while no smearing is applied on the $r$ coordinate. Optionally, an average hit inefficiency can be randomly applied to each detector layer; in this study, all layers are assumed to be fully efficient.

The signal targeted in this study is a sample of single tracks originating from a simulated hard scattering vertex with $p_T > 20$ GeV. Tracks are generated within the $|\eta|<1.4$ region to ensure full containment within the detector volume, thus avoiding acceptance effects on the hits. To emulate the typical beamspot shape in LHC running conditions, each track originates from a primary vertex (PV) with its position along the beam axis sampled from a Gaussian distribution centred at the origin, with a $\sigma$~=~5~cm. The longitudinal ($z_0$) impact parameter, evaluated with respect to the PV, is randomly sampled from a Gaussian distribution centred at 0 with $\sigma$~=~100~$\mu$m, while the transverse ($d_0$) impact parameter is set to 0.

Once an event containing a single signal track is simulated, two sources of background hits are overlaid. Pile-up collisions are simulated generating events with a variable number of tracks, following the longitudinal beamspot shape described above. To approximately replicate LHC Run~3 pile-up conditions, the $p_T$ distribution for pile-up tracks is tuned to match the measurements reported by ATLAS~\cite{ATLAS_charged_particles}. An average of 25 simultaneous collisions ($\langle\mu\rangle=25$) is overlaid on the signal sample. Additionally, random uncorrelated hits are added to each event to simulate detector noise, replacing what is typically observed in ATLAS-simulated samples.

Finally, to emulate the Region of Interest (RoI) data selection mechanism used in HLT tracking~\cite{ATLAS_trigger_tracking}, hits are preselected within a wedge with an opening of $|\Delta\eta|\times|\Delta\phi| = 0.1\times 0.1$ around the signal track, starting from $\pm 5$~mm around the PV. Only events that fully contain the signal track are considered to avoid potential acceptance effects in this preselection.

\begin{figure}[b]
    \centering
    \includegraphics[width=0.9\textwidth]{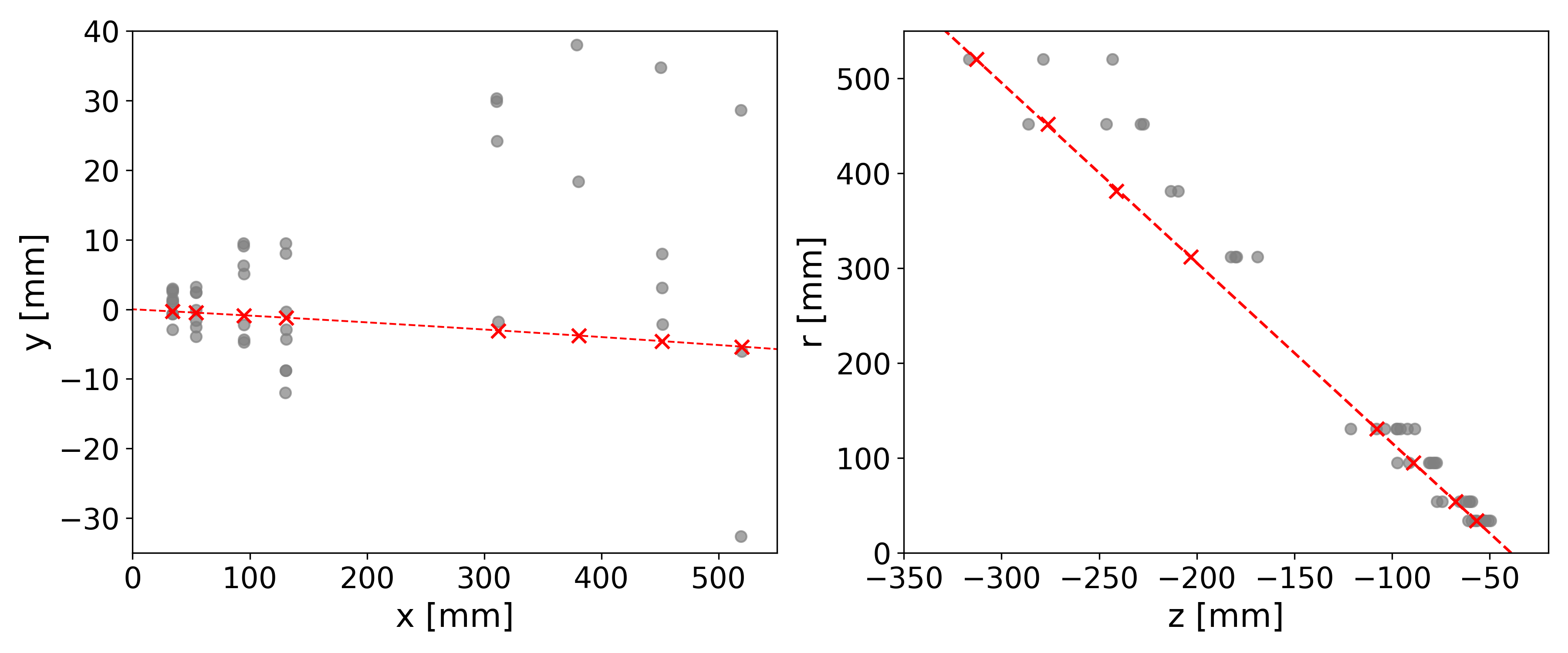}
    \caption{Display of a simulated event in the transverse (left) and longitudinal (right) views relative to the beam axis. Signal hits are shown in red, while pile-up and noise hits are displayed in grey. The dotted line represents the track, fitted as a circle in the $x-y$ plane and as a straight line in the $z-r$ plane.}
    \label{fig:event}
\end{figure}

An example event display of a generated signal track, along with its corresponding signal and background hits, is shown in Fig.~\ref{fig:event}.

The dataset used in this study consists of approximately 400k events for training and 50k events each for validation and testing.

\section{Network Architecture}
\begin{figure}[b]
    \centering
    \includegraphics[width=\textwidth]{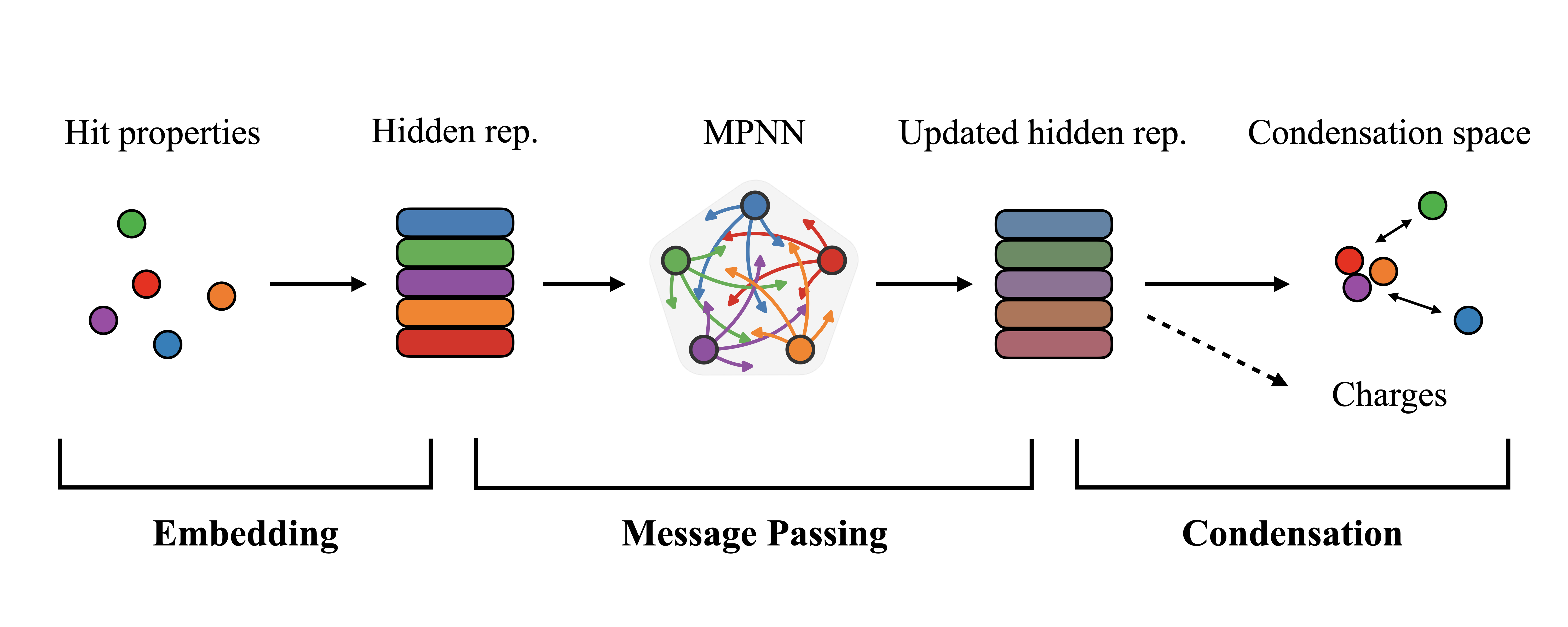}
    \caption{Sketch of the proposed network consisting of three blocks: embedding, message passing and condensation.}
    \label{fig:architecture}
\end{figure}
The task addressed by our GNN-based algorithm is to separate background hits from pile-up and noise from those produced by signal tracks and to distinguish between signal hits originating from different tracks.
We apply the condensation approach proposed for particle flow reconstruction~\cite{OC_jan,OC_2}, with a few adjustments. 

For demonstration purposes, the proposed network is intentionally designed to be small and compact. We adopt a graph-based method, where each graph node represents a hit, parametrized by its position in Cartesian coordinates, expressed in millimetres. The features are scaled to have a mean of 0 and a standard deviation of 1 across the entire dataset. For simplicity, the graphs are fully connected, though this can be generalized to larger datasets if the number of edges becomes incompatible with available memory. 

The nodes are then processed by the GNN, which consists of three blocks:
\begin{itemize}
    \item \textbf{Embedding.} For each node, the input features are transformed into a hidden representation of dimension 5 using a multilayer perceptron (MLP). This dimension was chosen to be as small as possible while maintaining performance, given the subsequent symbolic regression step. In real-world applications, a higher dimension may be necessary.
    \item \textbf{Message Passing (MP).} The hidden representations are passed through the edges as messages to all connecting nodes. The received messages are summed for each node, concatenated with its hidden representation, and processed through an MLP to update the representation.
    \item \textbf{Mapping to condensation space.} The updated hidden representations are mapped into a condensation space with a dimensionality of 5.
\end{itemize}
The network has a total of 446 k parameters, distributed approximately equally across the blocks. An illustration of the architecture is shown in Fig.~\ref{fig:architecture}. 

To cluster signal and background hits, the loss function consists of attractive and repulsive potentials, determined by a separately predicted charge for each node. The charges are calculated as follows: 
\begin{equation}
    q_i = \text{arctanh}(\beta_i)^2 + q_\text{min} 
\end{equation}
Where $i$ is the node index, $\beta$ is the output of a separate MLP using the updated hidden representation, and $q_\text{min}$ is a hyperparameter optimized to 0.4 via grid search. The loss for charge prediction is a Binary Cross Entropy (BCE) loss applied to $\beta$, classifying between signal and background hits. Ideally, background hits will have a charge of 0, while signal hits will have a high charge. 

Using this charge, we define the potentials for a given track $k$ by identifying the node with the maximal charge among all nodes associated with the track. This node is indexed as $\alpha$.

The attractive and repulsive potentials $\check{V}_k(x)$ and $\hat{V}_k(x)$ for track $k$, with $x$ representing the location in condensation space, are given by: 
\begin{align}
    \check{V}_k(x) &= ||x-x_\alpha||^2q_{\alpha k}\\\nonumber
    \hat{V}_k(x) &= \text{max}(0,1-||x-x_\alpha||)q_{\alpha k}
\end{align}
where $||\cdot||$ represents the L2 norm. The attractive potential grows quadratically with the distance from the node with the highest charge, while the repulsive potential decreases linearly until it reaches 0. 

The goal is to pull all nodes belonging to track $k$ towards the same point using the attractive potential while pushing all other nodes away using the repulsive potential. To achieve this, we define a matrix $M_{ik}$, where elements are set to 1 if node $i$ belongs to track $k$, and 0 otherwise, to define the potential loss $L_V$:
\begin{align}
    \begin{split}
    L_V =& \frac{1}{N}\sum_{j=1}^Nq_j\sum_{k=1}^K\left(M_{jk}\check{V}_k(x_j)+(1-M_{jk})\hat{V}_k(x_j)\right)  \\
    &\qquad+\frac{1}{N_{bkg}}\mathlarger{\sum}_{j=1}^{N_{bkg}}||x_j||\left(1-\sum_{k=1}^KM_{jk}\right)
    \end{split}
\end{align}
where $N$ is the total number of nodes in the event, and K is the total number of tracks. The first term in the loss, originally proposed in~\cite{OC_jan}, represents the combined force from all tracks acting on each node in the event, summed and normalized across all nodes.

The second term applies only to the $N_{bkg}$ nodes that do not belong to any signal track and originate from pile-up or noise. This term equals zero if the noise is located at the origin of the condensation space.
This setup simplifies postprocessing, as background removal can be easily achieved with a simple cut around the origin. The absence of signal hits in this area is enforced by the repulsive potential in the first loss term.
\section{Symbolic Regression}
The main goal of this work is to demonstrate that the above neural network can be approximated by a functional expression. For this purpose, we use the symbolic regression algorithm provided by the \textsc{PySR} package~\cite{pysr}. This algorithm is based on genetic programming, where trees of functions are randomly constructed using the provided operators and constraints. At each iteration, mutations are applied to the function trees, and the new trees are evaluated based on the specified loss function. If the performance of the new trees is better than that of the original ones, they are retained in the equation pool; otherwise, they are discarded. The output provides the best equations for all equation sizes, allowing the user to balance between accuracy and compactness.

An initial implementation of symbolic regression expressions on FPGAs was proposed in~\cite{fpga_sr} in the context of jet classification. There, a single MLP was used to process a set of jet variables. However, this approach is unsuitable for tracking tasks, as our input consists of set-valued data that cannot be processed by a simple MLP.

To address this challenge with a novel approach, we exploit the modularity of our model, which essentially consists of 3 MLPs with a minor processing step in between. We replace each MLP with a symbolic expression separately.\\
For the embedding network, the function operates uniformly on each node, taking the hit features as input and producing its hidden representation as output.\\
Given our current choice of a fully connected graph, the message-passing step simply sums the hidden representations of all hits. In future implementations, this may change, as an arbitrary graph would require differently masked sums for each hit, depending on its connectivity. Once handled, the MLP in the message-passing block again operates on each node individually, using the summed message as input. \\
For mapping into the condensation space, the function takes the updated hidden representation of each node and outputs its coordinates, similar to the first block.

We use default \textsc{PySR} parameters for the symbolic regression training. The selected operators include $+$, $-$, $\cdot$, $:$, square, cube, and ReLU(), with the maximal equation length set to 75 (150 for the final block). Prioritizing performance over compactness, we select the longest function from the output. Mean squared error (MSE) is used for function optimization.

Since \textsc{PySR} exhibits a significant decrease in speed for large datasets, we reduce the training dataset to 5,000 hits to optimize computational resources. In the adopted dataset, only 17\% of the hits are from signal tracks. To ensure \textsc{PySR} can correctly learn the signal features, we increase the ratio of signal to background hits to 50\% by downscaling the background hits randomly. 

We iteratively replace the network blocks with the learned symbolic expressions, starting with the embedding block. After each replacement, the remaining network is retrained to minimize performance loss. 

\section{Postprocessing}

The proposed GNN model is not designed to address the track reconstruction task fully; instead, it focuses on providing background hit filtering and track seeding information. To complete the track reconstruction and fitting, we implemented additional postprocessing steps. These steps are kept intentionally simple, as they are intended only to evaluate the quality of the seeding and the positive impact of background rejection on reducing the combinatorial complexity of the tracking problem.

\subsection{Hit selection}
In the original object condensation proposal, objects are identified based on the charge of the condensation node, with everything within a radius around this node being matched to the object. We modified this approach so that the evaluation does not rely on charge prediction, as this would require another symbolic replacement, introducing an additional source of error.

Due to the additional loss term, pile-up and noise hits are ideally clustered at the origin of the condensation space. Thus, we can eliminate background hits by applying a radius cut $r_{cut}$ around the origin.

Next, we apply the Men Shift (MS) clustering algorithm to identify signal hits. This clustering step will be critical for distinguishing hits from different tracks in multi-track datasets. For now, since each event contains only one signal track, the clustering primarily serves to remove background hits further. The clustering algorithm uses a bandwidth parameter that defines the size of the cluster. Along with the radius cut, these two parameters can be optimized. 

If the radius cut is too large, potential signal hits may be missed. If it is too small, high background contamination may remain, complicating further processing. A similar trade-off applies to the bandwidth parameter. To optimize background rejection ($r_b$) and signal efficiency  ($\varepsilon_s$), we perform a grid search over both parameters for each performance evaluation. Signal hit efficiency refers to the percentage of signal hits selected, while background rejection is the percentage of background hits removed.

\subsection{Track reconstruction and fit}
After the hit selection, we apply a simplified track reconstruction algorithm inspired by fast-tracking tasks developed by LHC experiments. To achieve this, we build all possible triplets from the selected hits. For each triplet, we construct a circle in the $x-y$ plane using the \textit{circle-fit} library to obtain the track-candidate $p_T$. We then perform a linear fit in the $r-z$ plane to extract $z_0$ using the non-linear least squares \textit{curve-fit} method from \textsc{SciPy}~\cite{scipy}.

With these results, we fill a 2D histogram in the $p_T-z_0$ plane and select triplets within a window around the histogram peak. The window size depends on the peak's $p_T$ value and is chosen large enough to encompass all hits if the triplets were built using true signal hits. Only hits from triplets within the selected window are retained, further improving background rejection. At this stage, we reevaluate the signal efficiency and background rejection.

If, after the above steps, we do not have more than one hit per layer, we obtain the final track parameters by performing separate fits in $r-z$ and $x-y$ as described above. If at least one layer contains more than one hit, we apply simplified ambiguity solving. In this case, we perform the linear $r-z$ fit for each combination and select the one with the best $\chi^2$. This process is repeated iteratively for each layer with more than one hit before performing the final circle fit.

We define hit purity as the percentage of signal hits selected for the final track fit.

After the fit, we examine the difference between the fitted values and true particle features. Due to finite detector resolution, even a fit on perfectly selected hits will not yield exact particle features. As a reference, we perform a \textit{truth fit}, in which we fit the track using the truth information of the signal hits. 

Furthermore, we aim to quantify the resolution of $p_T$. The expected resolution depends on $p_T$, as a high-energy track with little curvature is more sensitive to perturbations in hit positions than a low-energy track. To obtain a rough estimate of the ideal $p_T$ resolution, we divide the data into bins of true track $p_T$ and fit a Gaussian to the truth fit residuals for each bin, extracting $\sigma$. This allows us to measure the percentage of fits where the $p_T$ residual falls within the $1\sigma$ or $2\sigma$ range. 
\section{Results}
We present the results for the full GNN to demonstrate its potential as a tracking tool. This serves as a baseline for comparison with all the symbolic expression replacement options. We denote the network with the first stage replaced with SR~1, the one with both the first and second stages replaced with SR~2 and the fully approximated network with SR~3. For SR~1 and SR~2, the evaluations are conducted on the retrained network.
\begin{figure}[t]
    \centering
    \includegraphics[width=0.8\textwidth]{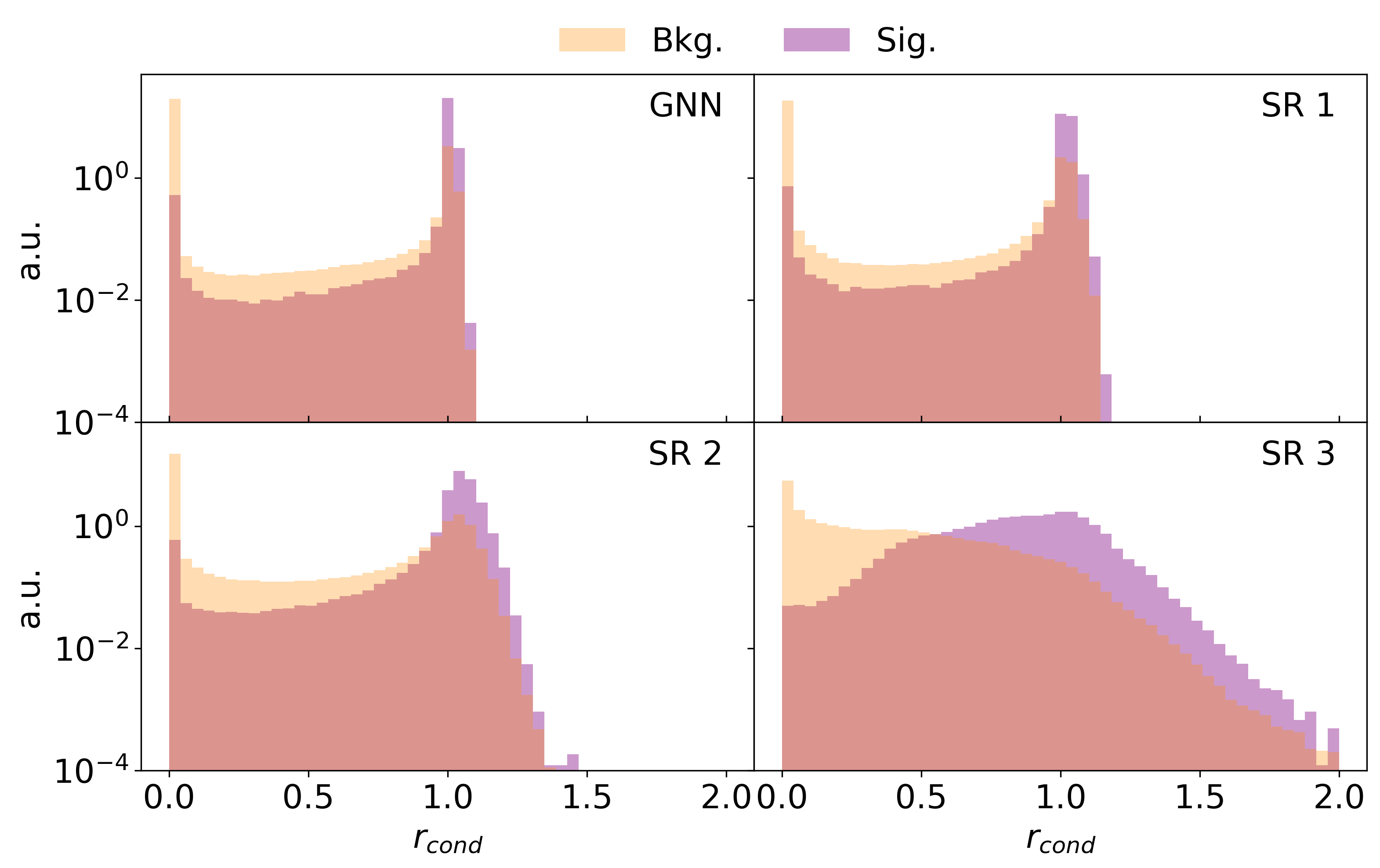}
    \caption{Condensation radius for all signal and background hits for the different stages.}
    \label{fig:radius}
\end{figure}
\begin{table}[b]
    \centering
    \begin{tabular}{lccccc}
    \toprule
         & Truth & GNN & SR~1 & SR~2 & SR~3 \\
         \midrule
       $r_{\text{cut}}$  & - & 0.05 & 0.2 &0.3 & 0.4 \\
       Bandwidth & - & 0.7 &  0.7 & 0.6 & 0.7 \\
       \midrule
       $\varepsilon_s$ cluster & - & 96.3\% & 94.8\% & 92.7\% & 85.5\% \\
       $r_b$ cluster & - & 93.9\% & 93.5\% & 93.0\% & 71.1\% \\
       $\varepsilon_s$ triplets & - & 95.0\% & 93.5\% & 91.3\% & 82.4\% \\
       $r_b$ triplets & - & 98.5\% & 98.4\% & 98.2\% & 90.1\% \\
       Hit purity & - & 96.8\% & 95.9\% & 94.9\% & 83.3\% \\
       \midrule 
       $p_T$ fit in $1\sigma$  & 39.8\% & 37.8\% & 37.5\% & 36.7\% & 29.7\% \\
       $p_T$ fit in $2\sigma$  & 97.6\% & 95.0\% & 94.4\% & 93.6\% & 84.5\% \\
       \bottomrule
    \end{tabular}
    \caption{Clustering parameters and performance metrics. The signal hit efficiencies ($\varepsilon_s$) and background rejections ($r_b$) are measured after the clustering and triplet filtering steps. Hit purity indicates the average percentage of signal hits used in the final best fit after resolving ambiguities among multiple hit candidates per layer. The track fit performance is assessed by comparing the fitted $p_T$ to the true track $p_T$, measuring how often their difference falls within 1 or 2 times the estimated ideal resolution ($\sigma$). "Truth" refers to the fit results obtained using only signal hits.}
    \label{tab:metrics}
\end{table}

First, we examine the pure condensation output. Fig.~\ref{fig:radius} shows the condensation radius for signal and background hits across the different models. The degradation in performance is evident with the widening of the peaks, indicating a drop in precision among the models.\\
As mentioned earlier, the cut-off radius $r_\text{cut}$ and clustering bandwidth are optimized via grid search for each model and summarized in Tab.~\ref{tab:metrics}. This demonstrates that achieving maximum signal efficiency in clustering requires loosening the $r_\text{cut}$. If the signal and background distributions are not sufficiently separated, a tight radius cut may leave background contamination too high, potentially leading to the identification of a large cluster without signal hits. To balance high background rejection and reasonable clustering, an increase in $r_\text{cut}$ becomes necessary as more stages are replaced.

Signal efficiencies and background rejections after clustering and triplet filtering, along with the hit purity after all processing steps, are detailed in Tab.~\ref{tab:metrics}. The triplet filtering step improves background rejection by about 5 \% while only reducing signal efficiency by roughly 1 \%. Comparing the models, we observe minimal performance degradation for SR~1 and SR~2 but a noticeable drop for SR~3. Hit purity declines by only 2 \% for SR~2 compared to the original GNN but decreases by over 13 \% for the full SR model. This is likely due to the lack of retraining possibilities after the last replacement stage. Similar performance drops are observed for the first network parts if no retraining is performed. Retraining helps recalibrate the SR outputs, recovering most of the performance losses.

\begin{figure}[b]
    \centering
    \begin{subfigure}{0.48\textwidth}
        \includegraphics[width=0.95\textwidth]{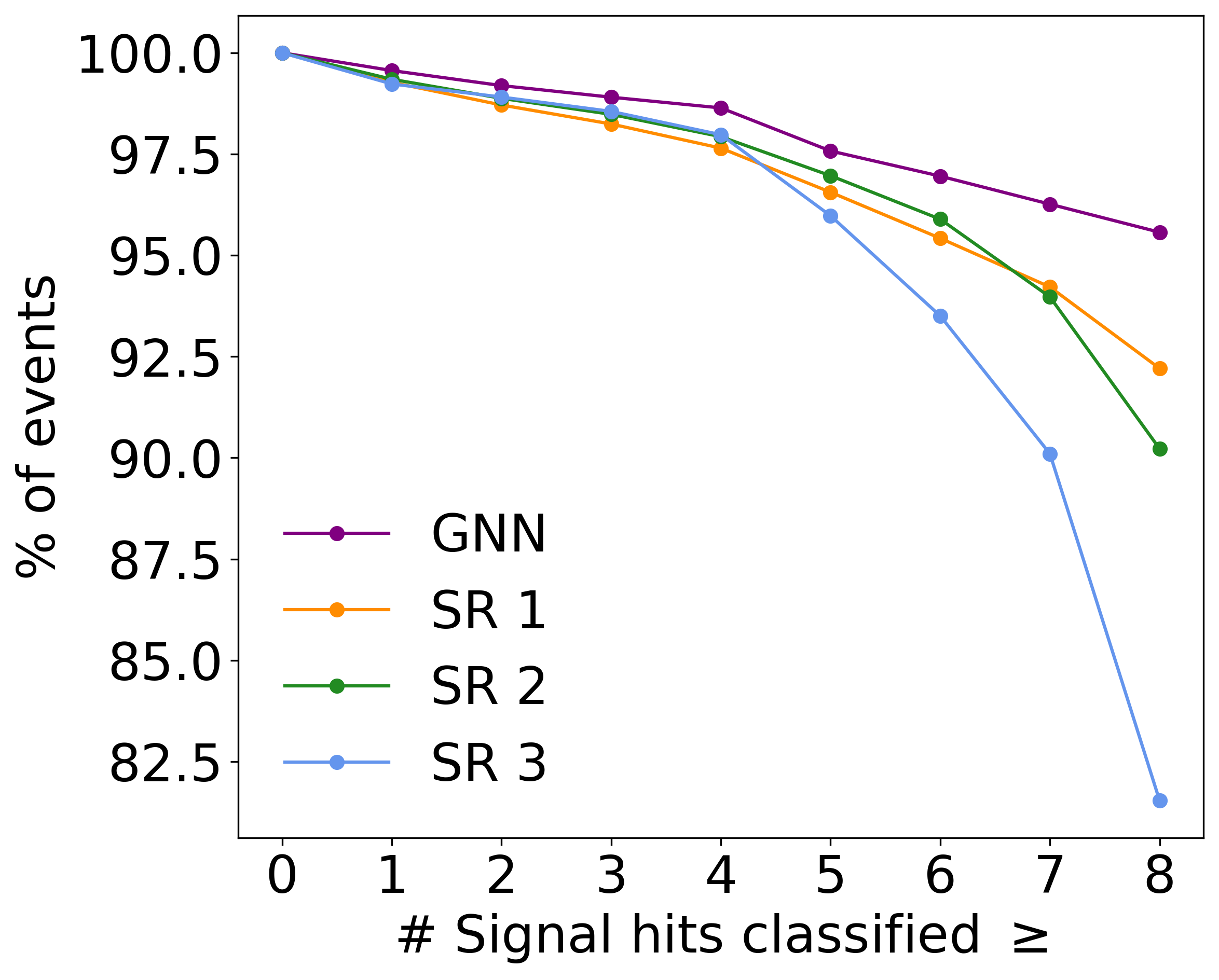}
    \end{subfigure}
    \begin{subfigure}{0.48\textwidth}
        \includegraphics[width=0.95\textwidth]{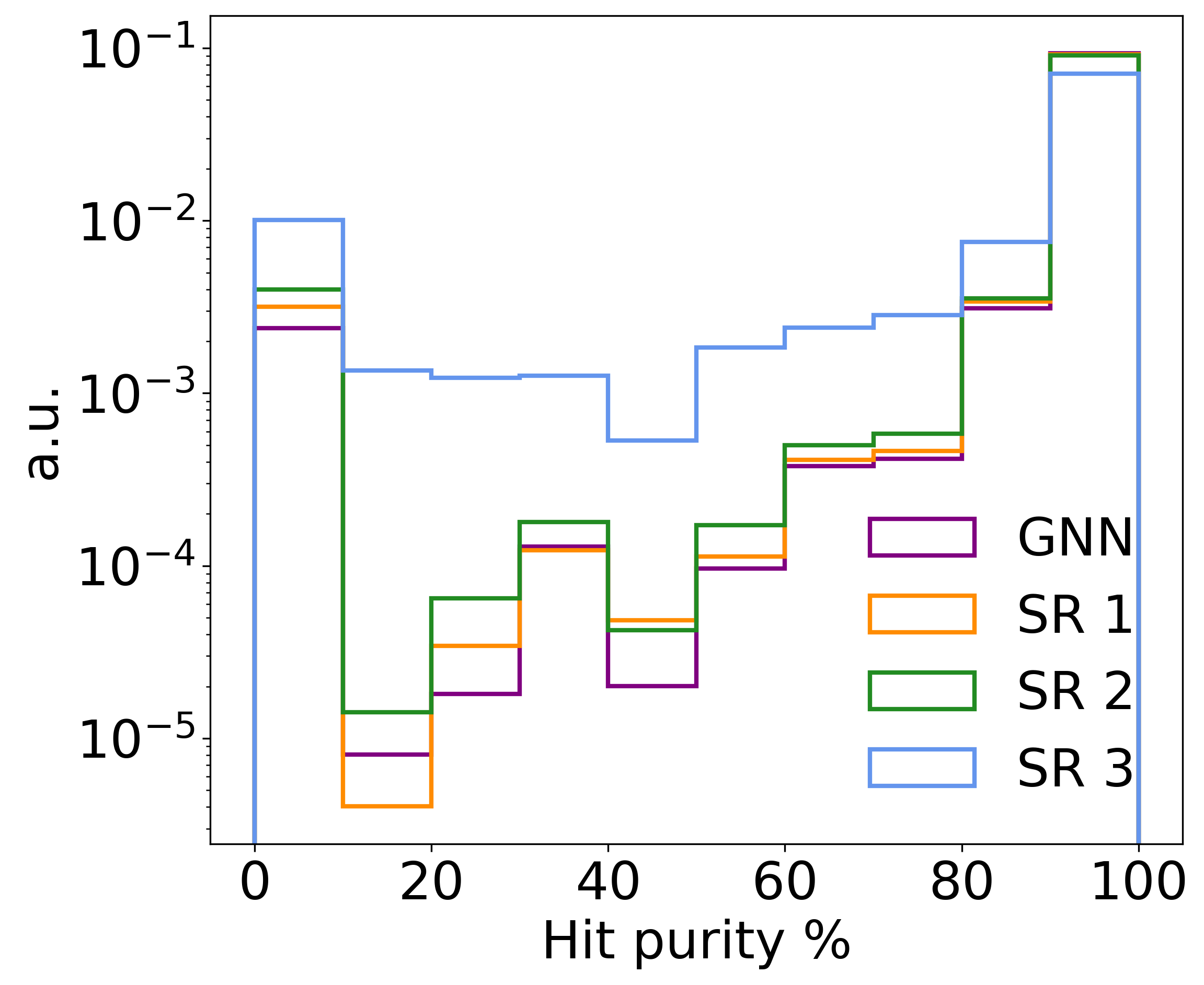}
    \end{subfigure}
    \caption{Percentage of events with $\geq N$ signal hits selected after clustering (left). Percentage of signal in the hits selected for the final track fit (right).}
    \label{fig:track}
\end{figure}

The left panel of Fig.~\ref{fig:track} shows the percentage of events retaining a given number of signal hits after clustering. The slight decrease in the probability of selecting all eight signal hits, compared to the individual signal hit efficiency reported in Tab.~\ref{tab:metrics}, highlights a strong correlation in signal misclassification. The network tends to either classify a full track correctly or misclassify the entire track. This trend is also evident in the final hit purity distribution, shown in the right panel of Fig.~\ref{fig:track}, where two distinct peaks appear at 0 and 100\%. For SR~1 and SR~2, these peaks remain sharp, whereas SR~3 shows a significant broadening.

\begin{figure}[t]
    \centering
    \begin{subfigure}{0.49\textwidth}
        \includegraphics[width=0.95\textwidth]{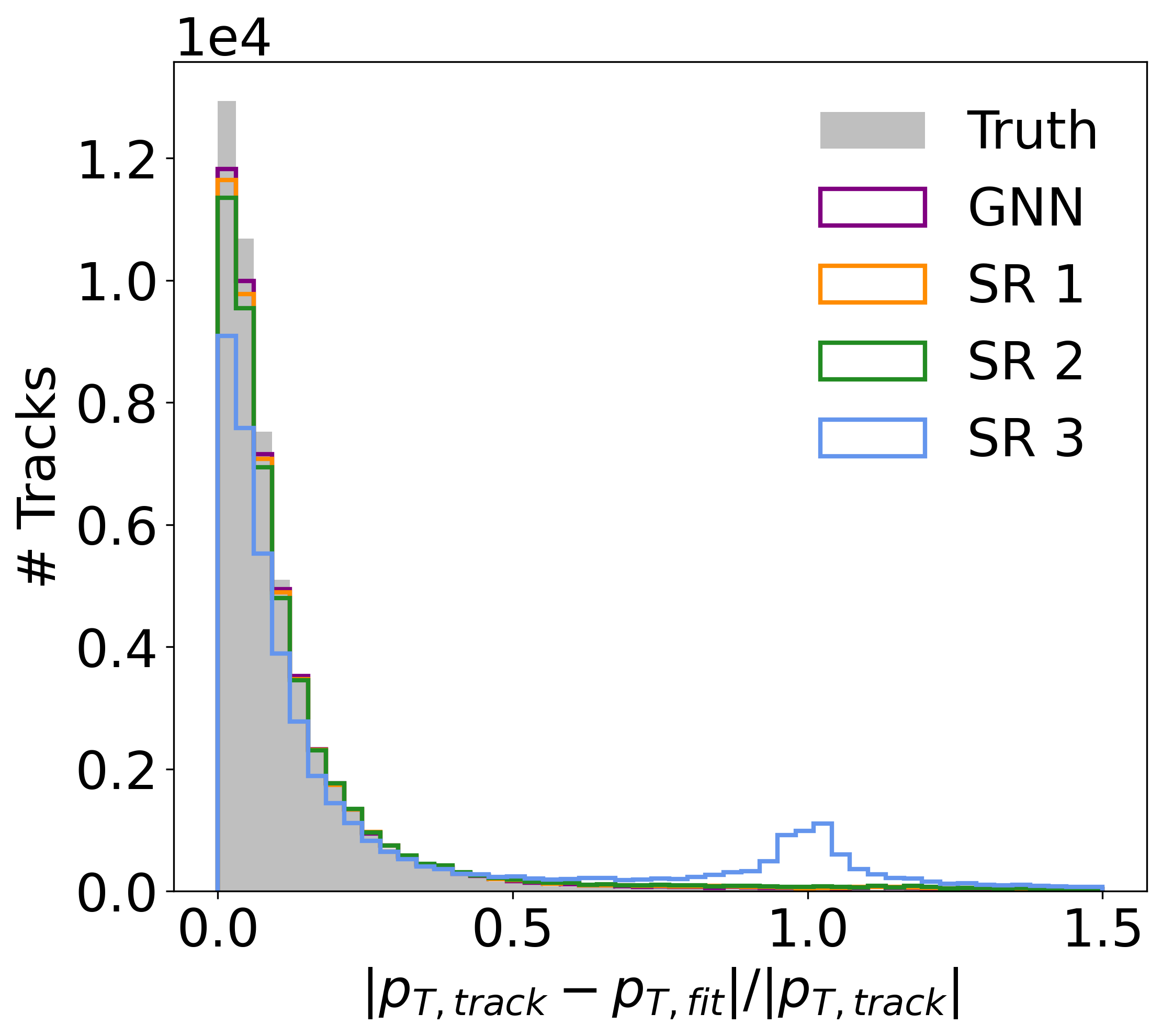}
    \end{subfigure}
    \begin{subfigure}{0.47\textwidth}
        \includegraphics[width=0.95\textwidth]{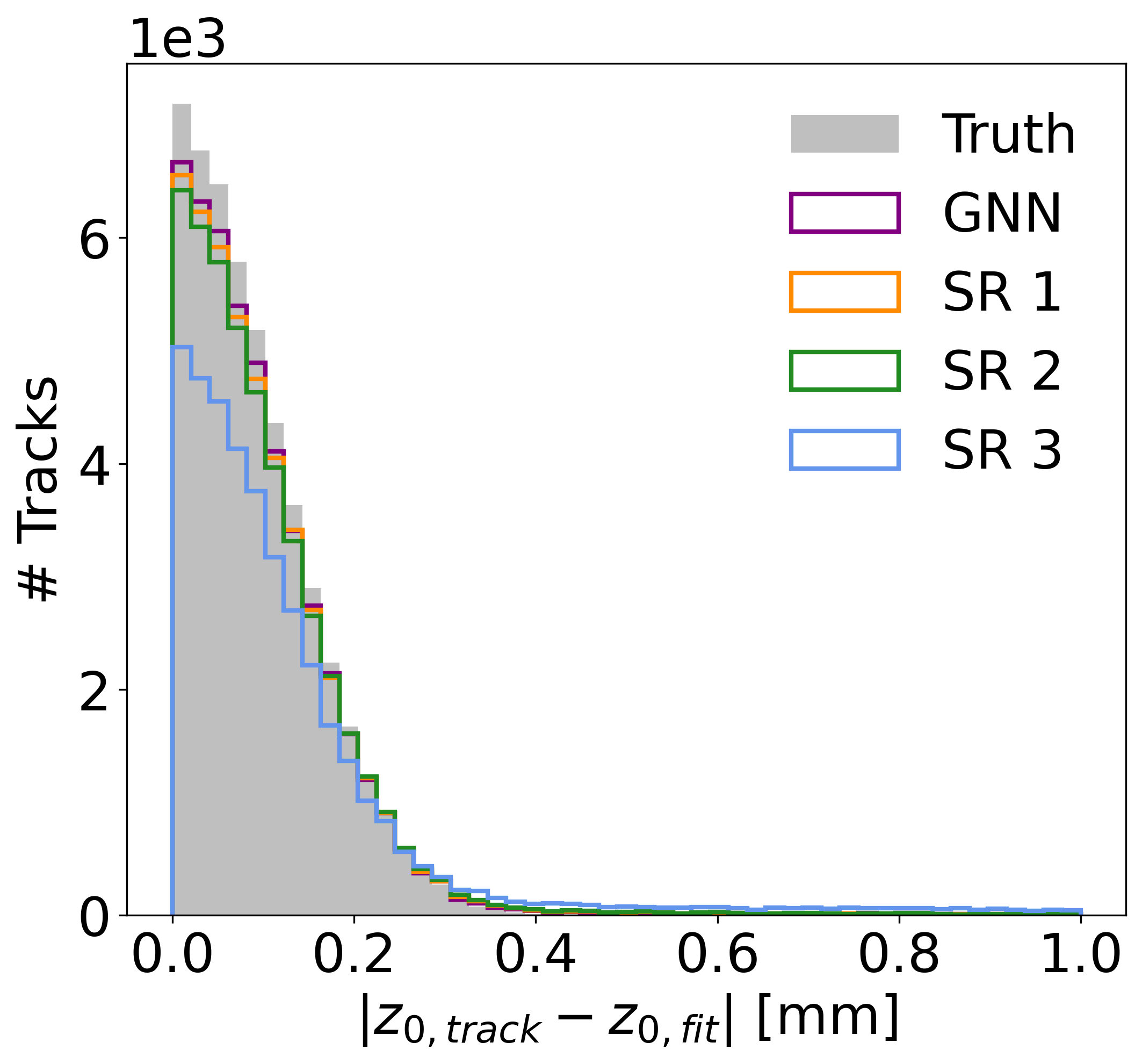}
    \end{subfigure}
    \caption{Residual of fitted track features. Normalized $p_T$ residual (left) and $z_0$ (right).}
    \label{fig:residual}
\end{figure}

The residuals between the true particle kinematics and the fitted $p_T$ and $z_0$ track parameters are shown in Fig.~\ref{fig:residual}, alongside the truth fit reference. As expected from the previous results, the performances for SR~1 and SR~2 show only minor deviations from the original GNN. When signal hit filtering fails completely, the circle fit tends to return a nearly null $p_T$ estimate, producing a small peak around 1 in the normalized residual distribution.

The $p_T$ resolution studies, shown at the bottom of Tab.~\ref{tab:metrics}, indicate similar performances across the models. We acknowledge certain limitations in the approach used to obtain these values, including the flexibility in choosing the $p_T$ binning, low statistics in the high-energy range, and the Gaussian fit's inadequacy in modelling the one-sided tails present in high $p_T$ regions. Despite these limitations, we believe the current approach is sufficient for the proof-of-concept demonstrated here.

Future work will focus on refining these methods, incorporating hit position uncertainties into the fit to extract more precise uncertainties on all fitted parameters. This will provide a more robust evaluation of tracking performance.

A potential application of the proposed network lies in its ability to eliminate background hits, significantly reducing the combinatorics involved in the tracking problem. The benefit of this can be measured by counting the number of triplets constructed after hit filtering for different models.

For a rough estimate, an ideal hit selection with only eight signal hits would construct 56 triplets. In the case of the full GNN, SR~1 and SR~2, we observe an average of approximately 100 triplets, while for SR~3, the average jumps to around 1000 triplets. Without any filtering, the number of triplets reaches approximately 13,000. This demonstrates that even with limited background rejection - about 70\% for the SR~3 - there is still more than a factor of 10 reduction in triplet processing time. 

This outcome highlights the trade-off between signal efficiency and background rejection and will be considered in future studies aimed at optimizing the network's performance, potentially prioritizing higher signal efficiency over stringent background rejection.

\section{Conclusion}
In this work, we introduced a new approach using symbolic regression to approximate complex neural networks for fast inference, with the goal of enabling feasible implementation on FPGAs. We began by breaking down a GNN into its three (MLP components, deriving analytical functions for each.

Although fast machine learning packages like \textsc{hls4ml}~\cite{hls4ml} can implement neural networks on FPGAs, their resources become constrained for networks exceeding 10k parameters. Symbolic functions, by contrast, offer a more resource-efficient solution. Our main objective was to demonstrate that using symbolic approximations for such networks in track-finding tasks within the trigger system does not result in significant performance losses.

Our findings show that replacing the first two MLP blocks with symbolic functions results in minimal performance loss, while replacing the final block leads to more significant performance degradation since retraining is not possible after the last stage, preventing output calibration. The flexibility of our method allows for the decision to either replace all blocks with symbolic regression or retain a small network in the final step to minimize performance loss.

Additionally, the purpose of the network can be adapted to different use cases. In this paper, we employed the network for track seeding, combining it with external postprocessing steps, such as clustering and final track-fitting. Alternatively, the network could serve as a pre-filtering step to reduce combinatorics and accelerate conventional track-finding algorithms. 

The feasibility of both approaches must be further evaluated on more complex and realistic datasets featuring multiple tracks. While the performance of symbolic regression on such datasets remains to be fully assessed, the more dispersed outputs produced by SR compared to the GNN could pose challenges for clustering in multi-track scenarios. However, increasing the network size and adding more dimensions may help address these challenges. It's important to note that while symbolic regression can easily handle a modest increase in dimensionality, scaling up significantly, such as by a factor of 100, would be less practical, as a symbolic expression has to be found for each dimension. Future work will also focus on implementing this method on FPGAs and conducting timing studies to evaluate its practical application. 

\section*{Acknowledgement}
NS and EG are supported by the
BSF-NSF Grant 2020780 and the ISF Grant 2871/19.
\newpage

\printbibliography

@article{ATLAS,
    author         = "{ATLAS Collaboration}",
    title          = "{The ATLAS Experiment at the CERN Large Hadron Collider}",
    journal        = "JINST",
    volume         = "3",
    year           = "2008",
    pages          = "S08003",
    doi            = "10.1088/1748-0221/3/08/S08003",
    primaryClass   = "hep-ex",
}

@article{CMS,
    author         = "{CMS Collaboration}",
    title          = "{The CMS Experiment at the CERN LHC}",
    journal        = "JINST",
    volume         = "3",
    year           = "2008",
    pages          = "S08004",
    doi            = "10.1088/1748-0221/3/08/S08004",
}

@article{LHC,
      author         = "Evans, Lyndon and Bryant, Philip",
      title          = "{LHC Machine}",
      journal        = "JINST",
      volume         = "3",
      pages          = "S08001",
      doi            = "10.1088/1748-0221/3/08/S08001",
      year           = "2008",
      SLACcitation   = "%%CITATION = JINST,3,S08001;%%",
}

@article{ATLAS_trigger_tracking,
    author         = "{ATLAS Collaboration}",
    title          = "{The ATLAS inner detector trigger performance in \(pp\) collisions at \(13\,\text{TeV}\) during LHC Run~2}",
    journal        = "Eur. Phys. J. C",
    volume         = "82",
    year           = "2022",
    pages          = "206",
    doi            = "10.1140/epjc/s10052-021-09920-0",
    reportNumber   = "CERN-EP-2021-076",
    eprint         = "2107.02485",
    archivePrefix  = "arXiv",
    primaryClass   = "hep-ex",
}

@article{CMS_trigger_gpu_tracking,
  
AUTHOR={Bocci, A. and Innocente, V. and Kortelainen, M. and Pantaleo, F. and Rovere, M.},   
	 
TITLE={Heterogeneous Reconstruction of Tracks and Primary Vertices With the CMS Pixel Tracker},      
	
JOURNAL={Frontiers in Big Data},      
	
VOLUME={3},           
	
YEAR={2020},      
	  
URL={https://www.frontiersin.org/articles/10.3389/fdata.2020.601728},       
	
DOI={10.3389/fdata.2020.601728},      
	
ISSN={2624-909X}
}

@article{ATLAS_charged_particles,
    author         = "{ATLAS Collaboration}",
    title          = "{Charged-particle distributions in \(\sqrt{s} = 13\,\text{TeV}\) \(pp\) interactions measured with the ATLAS detector at the LHC}",
    journal        = "Phys. Lett. B",
    volume         = "758",
    year           = "2016",
    pages          = "67",
    doi            = "10.1016/j.physletb.2016.04.050",
    reportNumber   = "CERN-EP-2016-014",
    eprint         = "1602.01633",
    archivePrefix  = "arXiv",
    primaryClass   = "hep-ex",
}

@article{OC_jan,
    author = "Kieseler, Jan",
    title = "{Object condensation: one-stage grid-free multi-object reconstruction in physics detectors, graph and image data}",
    eprint = "2002.03605",
    archivePrefix = "arXiv",
    primaryClass = "physics.data-an",
    doi = "10.1140/epjc/s10052-020-08461-2",
    journal = "Eur. Phys. J. C",
    volume = "80",
    number = "9",
    pages = "886",
    year = "2020"
}

@article{OC_2,
    author = "Qasim, Shah Rukh and Chernyavskaya, Nadezda and Kieseler, Jan and Long, Kenneth and Viazlo, Oleksandr and Pierini, Maurizio and Nawaz, Raheel",
    title = "{End-to-end multi-particle reconstruction in high occupancy imaging calorimeters with graph neural networks}",
    eprint = "2204.01681",
    archivePrefix = "arXiv",
    primaryClass = "physics.ins-det",
    doi = "10.1140/epjc/s10052-022-10665-7",
    journal = "Eur. Phys. J. C",
    volume = "82",
    number = "8",
    pages = "753",
    year = "2022"
}

@article{ATLAS_trigger,
    author = "ATLAS Collaboration",
    collaboration = "ATLAS",
    title = "{Operation of the ATLAS trigger system in Run 2}",
    eprint = "2007.12539",
    archivePrefix = "arXiv",
    primaryClass = "physics.ins-det",
    reportNumber = "CERN-EP-2020-109",
    doi = "10.1088/1748-0221/15/10/P10004",
    journal = "JINST",
    volume = "15",
    number = "10",
    pages = "P10004",
    year = "2020"
}

@article{ATLAS_trigger_run3,
    author = "ATLAS Collaboration",
    collaboration = "ATLAS",
    title = "{The ATLAS Trigger System for LHC Run 3 and Trigger performance in 2022}",
    eprint = "2401.06630",
    archivePrefix = "arXiv",
    primaryClass = "hep-ex",
    reportNumber = "CERN-EP-2023-299",
    month = "1",
    year = "2024"
}

@article{cms_trigger,
    author = "CMS Collaboration",
    collaboration = "CMS",
    title = "{The CMS trigger system}",
    eprint = "1609.02366",
    archivePrefix = "arXiv",
    primaryClass = "physics.ins-det",
    reportNumber = "CMS-TRG-12-001, CERN-EP-2016-160",
    doi = "10.1088/1748-0221/12/01/P01020",
    journal = "JINST",
    volume = "12",
    number = "01",
    pages = "P01020",
    year = "2017"
}

@techreport{cms_trigger_run3,
      author        = "Fontanesi, Elisa",
      collaboration = "CMS",
      title         = "{New trigger strategies for CMS during Run 3}",
      institution   = "CERN",
      reportNumber  = "CMS-CR-2022-256",
      address       = "Geneva",
      year          = "2022",
      url           = "https://cds.cern.ch/record/2842439",
      doi           = "10.22323/1.414.0681",
}

@article{HW-tracking,
    author = "Ryd, Anders and Skinnari, Louise",
    title = "{Tracking Triggers for the HL-LHC}",
    eprint = "2010.13557",
    archivePrefix = "arXiv",
    primaryClass = "physics.ins-det",
    doi = "10.1146/annurev-nucl-020420-093547",
    journal = "Ann. Rev. Nucl. Part. Sci.",
    volume = "70",
    pages = "171--195",
    year = "2020"
}

@misc{trackml-particle-identification,
    author = {Salzburger, Andreas and Germain, Cecile and Rousseau, David and Charpiat, Guillaume and Gray, Heather and Basara, Laurent and Hushchyn, Mikhail and Kiehn, Moritz and Calafiura, Paolo and Farrell, Steve},
    title = {TrackML Particle Tracking Challenge},
    publisher = {Kaggle},
    year = {2018},
    url = {https://kaggle.com/competitions/trackml-particle-identification}
}

@inbook{trackml-sum,
    author={Amrouche, S. and Basara, L. and Calafiura, P. and Estrade, V. and Farrell, S. and Ferreira, D. R. and Finnie, L. and Finnie, N. and Germain, C. and Gligorov, V. Vava and Golling, T. and Gorbunov, S. and Gray, H. and Guyon, I. and Hushchyn, M. and Innocente, V. and Kiehn, M. and Moyse, E. and Puget, J.-F. and Reina, Y. and Rousseau, D. and Salzburger, A. and Ustyuzhanin, A. and Vlimant, J. and Wind, J. S. and Xylouris, T. and Yilmaz, Y.},
    title = "{The Tracking Machine Learning challenge : Accuracy phase}",
    booktitle = "{The NeurIPS '18 Competition: From Machine Learning to Intelligent Conversations}",
    eprint = "1904.06778",
    archivePrefix = "arXiv",
    primaryClass = "hep-ex",
    doi = "10.1007/978-3-030-29135-8_9",
    month = "4",
    year = "2019"
}

@article{GNN_general,
    author = "Duarte, Javier and Vlimant, Jean-Roch",
    title = "{Graph Neural Networks for Particle Tracking and Reconstruction}",
    eprint = "2012.01249",
    archivePrefix = "arXiv",
    primaryClass = "hep-ph",
    doi = "10.1142/9789811234033_0012",
    month = "12",
    year = "2020"
}

@inproceedings{GNN_fpga,
    author={A. Heintz and V. Razavimaleki and J. Duarte and G. DeZoort and I. Ojalvo and S. Thais and M. Atkinson and M. Neubauer and L. Gray and S. Jindariani and N. Tran and P. Harris and D. Rankin and T. Aarrestad and V. Loncar and M. Pierini and S. Summers and J. Ngadiuba and M. Liu and E. Kreinar and Z. Wu},
    title = "{Accelerated Charged Particle Tracking with Graph Neural Networks on FPGAs}",
    booktitle = "{34th Conference on Neural Information Processing Systems}",
    eprint = "2012.01563",
    archivePrefix = "arXiv",
    primaryClass = "physics.ins-det",
    reportNumber = "FERMILAB-CONF-20-622-CMS-SCD",
    month = "11",
    year = "2020"
}

@article{gnn_fpga2,
    author={Elabd, A. and Razavimaleki, V. and Huang, S. and Duarte, J. and Atkinson, M. and DeZoort, G. and Elmer, P. and Hauck, S. and Hu, J. and Hsu, S. and Lai, B. and Neubauer, M. and Ojalvo, I. and Thais, S. and Trahms, M.},
    title = "{Graph Neural Networks for Charged Particle Tracking on FPGAs}",
    eprint = "2112.02048",
    archivePrefix = "arXiv",
    primaryClass = "physics.ins-det",
    doi = "10.3389/fdata.2022.828666",
    journal = "Front. Big Data",
    volume = "5",
    pages = "828666",
    year = "2022"
}

@article{gnn_hllhc,
    author = "Biscarat, Catherine and Caillou, Sylvain and Rougier, Charline and Stark, Jan and Zahreddine, Jad",
    title = "{Towards a realistic track reconstruction algorithm based on graph neural networks for the HL-LHC}",
    eprint = "2103.00916",
    archivePrefix = "arXiv",
    primaryClass = "physics.ins-det",
    doi = "10.1051/epjconf/202125103047",
    journal = "EPJ Web Conf.",
    volume = "251",
    pages = "03047",
    year = "2021"
}

@inproceedings{gnn_Thais,
    author = "Thais, Savannah and DeZoort, Gage",
    title = "{Instance Segmentation GNNs for One-Shot Conformal Tracking at the LHC}",
    booktitle = "{34th Conference on Neural Information Processing Systems}",
    eprint = "2103.06509",
    archivePrefix = "arXiv",
    primaryClass = "cs.CV",
    month = "3",
    year = "2021"
}

@article{gnn_edgeclass,
    author = "DeZoort, Gage and Thais, Savannah and Duarte, Javier and Razavimaleki, Vesal and Atkinson, Markus and Ojalvo, Isobel and Neubauer, Mark and Elmer, Peter",
    title = "{Charged Particle Tracking via Edge-Classifying Interaction Networks}",
    eprint = "2103.16701",
    archivePrefix = "arXiv",
    primaryClass = "hep-ex",
    doi = "10.1007/s41781-021-00073-z",
    journal = "Comput. Softw. Big Sci.",
    volume = "5",
    number = "1",
    pages = "26",
    year = "2021"
}

@inproceedings{gnn_equivariant,
    author = "Murnane, Daniel and Thais, Savannah and Thete, Ameya",
    title = "{Equivariant Graph Neural Networks for Charged Particle Tracking}",
    booktitle = "{21th International Workshop on Advanced Computing and Analysis Techniques in Physics Research}: {AI meets Reality}",
    eprint = "2304.05293",
    archivePrefix = "arXiv",
    primaryClass = "physics.ins-det",
    month = "4",
    year = "2023"
}

@inproceedings{gnn_hierarchical,
    author = "Liu, Ryan and Calafiura, Paolo and Farrell, Steven and Ju, Xiangyang and Murnane, Daniel Thomas and Pham, Tuan Minh",
    title = "{Hierarchical Graph Neural Networks for Particle Track Reconstruction}",
    booktitle = "{21th International Workshop on Advanced Computing and Analysis Techniques in Physics Research}: {AI meets Reality}",
    eprint = "2303.01640",
    archivePrefix = "arXiv",
    primaryClass = "hep-ex",
    month = "3",
    year = "2023"
}

@inproceedings{gnn_ckf,
    author = "Heinrich, Lukas and Huth, Benjamin and Salzburger, Andreas and Wettig, Tilo",
    title = "{Combined track finding with GNN \& CKF}",
    eprint = "2401.16016",
    archivePrefix = "arXiv",
    primaryClass = "hep-ex",
    month = "1",
    year = "2024"
}

@article{fast_muon_fpga,
    author = "Sun, Chang and Nakajima, Takumi and Mitsumori, Yuki and Horii, Yasuyuki and Tomoto, Makoto",
    title = "{Fast muon tracking with machine learning implemented in FPGA}",
    eprint = "2202.04976",
    archivePrefix = "arXiv",
    primaryClass = "physics.ins-det",
    doi = "10.1016/j.nima.2022.167546",
    journal = "Nucl. Instrum. Meth. A",
    volume = "1045",
    pages = "167546",
    year = "2023"
}

@article{fpga_generic,
    author={Duarte, J. and Han, S. and Harris, P. and Jindariani, S. and Kreinar, E. and Kreis, B. and Ngadiuba, J. and Pierini, M. and Rivera, R. and Tran, N. and Wu, Z.},
    title = "{Fast inference of deep neural networks in FPGAs for particle physics}",
    eprint = "1804.06913",
    archivePrefix = "arXiv",
    primaryClass = "physics.ins-det",
    reportNumber = "FERMILAB-PUB-18-089-E",
    doi = "10.1088/1748-0221/13/07/P07027",
    journal = "JINST",
    volume = "13",
    number = "07",
    pages = "P07027",
    year = "2018"
}

@article{fpga_hls4ml,
    author={Ngadiuba, J. and Loncar, V. and Pierini, M. and Summers, S. and Di Guglielmo, G. and Duarte, J. and Harris, P. and Rankin, D. and Jindariani, S. and Liu, M. and Pedro, K. and Tran, N. and Kreinar, E. and Sagear, S. and Wu, Z. and Hoang, D.},
    title = "{Compressing deep neural networks on FPGAs to binary and ternary precision with HLS4ML}",
    eprint = "2003.06308",
    archivePrefix = "arXiv",
    primaryClass = "cs.LG",
    reportNumber = "FERMILAB-PUB-20-167-PPD-SCD, FERMILAB-PUB-20-167-PPD-SCD",
    doi = "10.1088/2632-2153/aba042",
    journal = "Mach. Learn. Sci. Tech.",
    volume = "2",
    pages = "015001",
    year = "2021"
}

@software{hls4ml,
author       = {{FastML Team}},
title        = {fastmachinelearning/hls4ml},
year         = 2023,
publisher    = {Zenodo},
version      = {v0.8.1},
doi          = {10.5281/zenodo.1201549},
url          = {https://github.com/fastmachinelearning/hls4ml}
}

@article{fpga_traf,
    author = "Jiang, Zhixing and Yin, Dennis and Khoda, Elham E. and Loncar, Vladimir and Govorkova, Ekaterina and Moreno, Eric and Harris, Philip and Hauck, Scott and Hsu, Shih-Chieh",
    title = "{Ultra Fast Transformers on FPGAs for Particle Physics Experiments}",
    eprint = "2402.01047",
    archivePrefix = "arXiv",
    primaryClass = "cs.LG",
    month = "2",
    year = "2024"
}

@article{gnn_fpga_reco,
    author={Iiyama, Y. and Cerminara, G. and Gupta, A. and Kieseler, J. and Loncar, V. and Pierini, M. and Qasim, S. R. and Rieger, M. and Summers, S. and Van Onsem, G. and Wozniak, K. A. and Ngadiuba, J. and Di Guglielmo, G. and Duarte, J. and Harris, P. and Rankin, D. and Jindariani, S. and Liu, M. and Pedro, K. and Tran, N. and Kreinar, E. and Wu, Z.},
    title = "{Distance-Weighted Graph Neural Networks on FPGAs for Real-Time Particle Reconstruction in High Energy Physics}",
    eprint = "2008.03601",
    archivePrefix = "arXiv",
    primaryClass = "physics.ins-det",
    reportNumber = "FERMILAB-PUB-20-405-E-SCD",
    doi = "10.3389/fdata.2020.598927",
    journal = "Front. Big Data",
    volume = "3",
    pages = "598927",
    year = "2020"
}

@article{fpga_convnet,
    author={Aarrestad, T. and Loncar, V. and Ghielmetti, N. and Pierini, M. and Summers, S. and Ngadiuba, J. and Petersson, C. and Linander, H. and Iiyama, Y. and Di Guglielmo, G. and Duarte, J. and Harris, P. and Rankin, D. and Jindariani, S. and Pedro, K. and Tran, N. and Liu, M. and Kreinar, E. and Wu, Z. and Hoang, D.},
    title = "{Fast convolutional neural networks on FPGAs with hls4ml}",
    eprint = "2101.05108",
    archivePrefix = "arXiv",
    primaryClass = "cs.LG",
    reportNumber = "FERMILAB-PUB-21-130-SCD",
    doi = "10.1088/2632-2153/ac0ea1",
    journal = "Mach. Learn. Sci. Tech.",
    volume = "2",
    number = "4",
    pages = "045015",
    year = "2021"
}

@article{SR_first,
    author = "Butter, Anja and Plehn, Tilman and Soybelman, Nathalie and Brehmer, Johann",
    title = "{Back to the formula - LHC edition}",
    eprint = "2109.10414",
    archivePrefix = "arXiv",
    primaryClass = "hep-ph",
    doi = "10.21468/SciPostPhys.16.1.037",
    journal = "SciPost Phys.",
    volume = "16",
    number = "1",
    pages = "037",
    year = "2024"
}

@article{SR_BSM,
    author = "AbdusSalam, Shehu and Abel, Steven and Crispim Rom\~ao, Miguel",
    title = "{Symbolic Regression for Beyond the Standard Model Physics}",
    eprint = "2405.18471",
    archivePrefix = "arXiv",
    primaryClass = "hep-ph",
    reportNumber = "IPPP/24/27",
    month = "5",
    year = "2024"
}

@article{fpga_sr,
    author = "Tsoi, Ho Fung and Pol, Adrian Alan and Loncar, Vladimir and Govorkova, Ekaterina and Cranmer, Miles and Dasu, Sridhara and Elmer, Peter and Harris, Philip and Ojalvo, Isobel and Pierini, Maurizio",
    title = "{Symbolic Regression on FPGAs for Fast Machine Learning Inference}",
    eprint = "2305.04099",
    archivePrefix = "arXiv",
    primaryClass = "cs.LG",
    doi = "10.1051/epjconf/202429509036",
    journal = "EPJ Web Conf.",
    volume = "295",
    pages = "09036",
    year = "2024"
}

@misc{pysr,
      title={Interpretable Machine Learning for Science with PySR and SymbolicRegression.jl}, 
      author={Miles Cranmer},
      year={2023},
      eprint={2305.01582},
      archivePrefix={arXiv},
      primaryClass={astro-ph.IM}
}

@article{SR_HI,
    author = "Mengel, Tanner and Steffanic, Patrick and Hughes, Charles and da Silva, Antonio Carlos Oliveira and Nattrass, Christine",
    title = "{Interpretable machine learning methods applied to jet background subtraction in heavy-ion collisions}",
    eprint = "2303.08275",
    archivePrefix = "arXiv",
    primaryClass = "hep-ex",
    doi = "10.1103/PhysRevC.108.L021901",
    journal = "Phys. Rev. C",
    volume = "108",
    number = "2",
    pages = "L021901",
    year = "2023"
}

@ARTICLE{scipy,
  author  = {SciPy Team},
  title   = {{{SciPy} 1.0: Fundamental Algorithms for Scientific
            Computing in Python}},
  journal = {Nature Methods},
  year    = {2020},
  volume  = {17},
  pages   = {261--272},
  adsurl  = {https://rdcu.be/b08Wh},
  doi     = {10.1038/s41592-019-0686-2},
}
\end{document}